# The Bell inequality is satisfied by quantum correlations computed consistently with quantum non-commutation

Louis Sica
lousica@jhu.edu
Institute for Quantum Studies, Chapman University, Orange, CA & Burtonsville, MD, USA
Inspire Institute Inc., Alexandria, VA, USA

**Abstract**

In constructing his theorem, Bell assumed that correlation functions among non-commuting variables are the same as those among commuting variables. However, in quantum mechanics, multiple data values exist simultaneously for commuting operations while for non-commuting operations data are conditional on prior outcomes, or may be predicted as alternative outcomes of the non-commuting operations. Given these qualitative differences, there is no reason why correlation functions among non-commuting variables should be the same as those among commuting variables, as assumed by Bell. When data for commuting and noncommuting operations are predicted from quantum mechanics, their correlations are different, and they now satisfy the Bell inequality.

**Keywords**
Bell's theorem; Bell inequality; hidden variables; correlations; commutation; noncommutation

**1. Introduction**

**1.1 Purpose**

Recently [1], Sica pointed out explicit logical difficulties resulting from the use of counterfactuals of non-commuting operations in the reasoning of the GHZ (Greenberger, Horne, Zeilinger) and Bell theorems. Detailed analysis of the failure of commutation in the GHZ theorem was given, and the implication of combining counterfactuals of non-commuting operations on the logic of the Bell theorem was outlined. The purpose of the present paper is to fill in that outline and correct logical inconsistencies in the use of counterfactuals in the Bell theorem.

It is found that the paradoxical violation of the Bell inequality by quantum mechanical correlations stems from ignoring the implication of non-commutation among the variables used (see Figure 1). This may have been based on non-commutation being regarded as a unique idiosyncratic property of quantum mechanics, which separates it from the classical world. It is also consistent with an almost total lack of recognition of non-commutation as a condition in the classical/macroscopic world. However, upon reflection, it will be recognized that non-commutative operations are encountered fairly frequently in everyday life. They are simply executed in an appropriate order, without contemplation of underlying logical principles. This reaches the point of absurdity in the expression "putting on shoes and socks," where the appropriate sequence of actions is executed rather than the sequence suggested by the language.

The purpose of the forgoing discussion is to explain why a critically important component in the logic of the Bell theorem has remained unrecognized. That component is the following: when non-commuting operations A (putting on shoes) and B (putting on socks) are considered separately, without regard to order, they are exclusive-OR statements that cannot be combined. The combined statement "carry out A alone and carry out B alone" makes no sense unless one is comparing different alternatives: e.g., wearing shoes alone on Monday, and wearing socks alone on Tuesday. However, if the counterfactuals (predicted results without performance) of non-commutative operations A and B are combined, other than as alternatives, their order of execution must be taken into account. Adhering to these logical principles, correlations of the variables used by Bell may be computed using quantum mechanical probabilities to obtain three simultaneously cross-correlated data sets. These cross-correlations satisfy the Bell inequality but are not second order stationary (SOS) in the way that Bell assumed because some measurement operations are non-commutative. (A random function $A(x)$ of time or space coordinates that results from a random process is defined to be SOS if $E\{A(x_1)\} = const.$ and $E\{A(x_1)A^*(x_2)\} = R(x_1 - x_2)$. Averages are denoted by $E\{\}$.)

Bell's theorem and the violation of Bell inequalities by correlations of experimentally measured data have been the subject of extensive discussion since the publication of Bell's historic 1964 paper [2], and of experimental results such as those of Aspect [3] and Weihs [4]. Although it is widely believed that violation of Bell inequalities implies non-locality, or the impossibility of quantum mechanical hidden variables, a number of authors have voiced objections to this opinion.

One of the earliest objections to Bell's work was given by de La Peña [5] who pointed out that Bell neglected the non-commutation of some of the variables used in his theorem. Bell explicitly responded to this objection [6], and pointed out that the variables in question were simply the predicted results of unperformed measurements at alternative instrument settings (counterfactuals). Unfortunately, Bell did not explicitly compute values of these counterfactuals using quantum mechanics, but assumed that if they could be accounted for by a random process underlying quantum mechanics, it must be a second order stationary (SOS) process [7]. A number of authors have voiced objections to aspects of Bell's logic in the construction of his theorem (of which a sampling is given in [8]-[11]). However, the physics community overall has accepted it.

## 1.2 Plan of this paper

To exhibit the mathematical facts as clearly as possible, in Section (2) the Bell inequality is first derived as a mathematical restriction on the simultaneous cross-correlations (SCC) of data sets, independently of their physical or probabilistic attributes. Although similar derivations have been given previously by workers in the probability area (see Hess [9] with references back to Boole [12] in 1862), and by Eberhard [13], certain logical implications of the basic mathematical facts with regard to conclusions of the Bell theorem have not previously been realized, as far as is known to the author. (In the supplementary Appendix, Bell's derivation is reviewed to examine the assumptions used beyond those he stated.)

In Section (3) a hidden variable algorithm is applied to a closely related situation to that which Bell considered. It produces Bell cosine correlations between a measurement and each of two alternative measurement counterfactuals, all occurring on one particle. However, a different correlation is produced between the non-commuting counterfactuals than between the realizable (though non-commuting) measurements. The same algorithm may be applied to two Bell particles using nonlocal information, or as here, to predict alternative counterfactual measurements on one particle using local information. The correlations from both situations satisfy the Bell inequality, but would violate it under Bell's assumption of a (SOS) process.

## 2. Derivation of the Bell inequality

The Bell inequality is widely believed to result from Bell's assumptions of hidden variables and locality. However, the inequality itself may be viewed as a mathematical fact independently of these assumptions; it results from the *procedure* of simultaneous cross-correlation (SCC) of three finite data sets. To make this clear, it will be derived using a simpler notation than that used by Bell. Its applicability to the situation that concerned Bell may then be examined.

Assume three lists, $a$, $b$, and $b'$ of length $N$, composed of elements $a_i$, $b_i$, $b_i'$ each of which equals $\pm 1$. From the ith elements of the three lists form

$$a_i b_i - a_i b_i' = a_i(b_i - b_i') = a_i b_i (1 - b_i b_i').$$  (2.1)

Summing over the $N$ elements of the lists, dividing by $N$, and taking absolute values on both sides, yields

$$\left| \sum_{i=1}^{N} a_i b_i / N - \sum_{i=1}^{N} a_i b_i' / N \right| \leq \sum_{i}^{N} |a_i b_i| |1 - b_i b_i'| / N = \sum_{i}^{N} (1 - b_i b_i') / N.$$  (2.2)

Upon further simplification of the right side, one obtains the Bell inequality for finite data sets:

$$\left|\sum_{i=1}^{N} a_i b_i / N - \sum_{i=1}^{N} a_i b_i^{'} / N\right| \leq 1 - \sum_{i}^{N} b_i b_i^{'} / N \quad . \tag{2.3}$$

The essence of SCC as defined in Equation (2.1) resides in the fact that the value of $a_i$ that multiplies $b_i$ is the same as that which multiplies $b_i^{'}$ for each $i$, and that the values of $b_i$ and $b_i^{'}$ for each $i$ are the same on both sides of Equations (2.1-2.3), i.e., the rules of algebra hold. Note: *this fundamental requirement for validity of the inequality is not satisfied when pairs of random variables are measured in statistically independent runs but will be satisfied by explicitly calculated counterfactual data.*

If limits exist for the averages in (2.3) as $N \rightarrow \infty$,

$$\left|\langle ab \rangle - \langle ab' \rangle\right| \leq 1 - \langle bb' \rangle , \tag{2.4}$$

where the symbol $\langle \; \rangle$ is used to indicate assumed limits in the case of averages over infinite data sets. For random processes using real data, such limits are commonly observed to hold, and will be assumed to hold here. (Note that inequalities (2.3-2.4) still hold if the absolute value sign is removed.)

A first crucial conclusion follows. In the case of experimental random process data for which $N$ is necessarily finite, fluctuations about the ensemble averages do not impair the validity of the Inequality (2.3). Simultaneous numerical cross-correlation values in Inequality (2.3) fluctuate together no matter how quantitatively limited the data might be. Even systematic errors have no effect on Inequality (2.3) as long as there is simultaneous cross-correlation over sets of data triplets.

A second conclusion is that although Inequality (2.3) is identically satisfied by SCC of any three data sets whatsoever (consisting of $\pm 1's$), Inequality (2.4) is not necessarily satisfied by any three assumed correlation functions *unless those functions result from SCC of data triplets*. The values of possible simultaneous cross-correlations are restricted by Inequality (2.4). Note that starting from data, it is very easy to satisfy the Bell inequality; even made-up data will suffice. What one cannot do, however, is make up the resulting correlations among the data sets, for if proposed sets of correlations violate the inequality, no data exist that can produce them under SCC.

The foregoing logic is relevant to current practice in Bell experiments. If the quantum mechanical measurements to which the Bell inequality has been applied could be represented by a SOS process, the data sets would not have to be simultaneously cross-correlated to obtain their correlation function. Since only one functional form would exist, it would not matter whether it was observed in simultaneous cross-correlation of data from different variables' settings or from independent runs. The function would satisfy the Bell inequality *except* for noise fluctuations.

Finally, the application of the Bell inequality to the Bell experiment schematic shown in Fig. 1 must be considered. A commonly proposed evasion of the simple logic leading to Inequality (2.3) has rested on the *ab initio* assumption of non-locality [14]. If the readout on the *A*-side of the apparatus $A(\theta_a)$ is replaced by $A(\theta_a, \theta_b)$ due to its dependence on the *B*-side setting, then a change in *B*-setting to $\theta_{b'}$, corresponding to an alternative counterfactual measurement, results in a change in the *A*-side outcome to $A(\theta_a, \theta_{b'})$. Equation (2.1) now no longer holds since the value of $a_i$ is changed when multiplied by $b_i^{'}$. This difficulty may be avoided, however, if the *A*-measurement is completed before the second detector angle is chosen, by using a longer path on the *B*-side and waiting the necessary increment of time for the *B*-particle to arrive. (Unequal optical path experiments have been performed [15].) After the *A*-measurement occurs, any assumed non-local influences now travel from *A* to *B*. Three-number data sets result from two real measurements *A* and *B*, and predicted counterfactual *B'*, so that the Bell Inequality (2.3) as derived applies. Such an experiment could test for evidence of non-local effects by using different settings for *B* before and after an event is registered at *A*.

## 3. Quantum correlations among two measurements and a counterfactual

When comparing counterfactuals of alternative non-commuting operations, one must compare predicted measurement outcomes at alternate settings, since macroscopic measuring devices cannot exist simultaneously at more than one setting. If measurements of *A* and *B* commute, as in conventional Bell experiments (see Figure 1) they may in principle be measured simultaneously, since both results exist at the same time. If two measurement operations do not commute, their order of execution determines the outcomes. However, one may still *compare predicted alternative values B and B'* (couterfactuals) from sequences $A-B$ and $A-B'$. Using conditional probabilities based on non-commutation, the correlation function $\langle BB'\rangle$ may be computed, but will in general be different from that of $\langle AB\rangle$ and $\langle AB'\rangle$. In the present situation, the values of all variables are $\pm 1$.

This is the situation considered by Bell in the construction of his theorem. It has been treated in [8], but a variation is considered here in which one real and two alternate counterfactual measurements are carried out on one particle. This eliminates the need for nonlocal information. The same correlations are produced as in the conventional Bell case [8] except for a minus sign that does not effect Inequality (2.4). The relevant conditional probabilities in this situation are [16]

$$P(B=\pm 1|A=\pm 1)=\cos^2\frac{\theta_B-\theta_A}{2}; \; P(B=\mp 1|A=\pm 1)=\sin^2\frac{\theta_B-\theta_A}{2}, \qquad (3.6)$$

where measurement *B* occurs after *A* on the same particle. Measurement values of *A* and *B* are inferred from the path through two Stern-Gerlach magnets and deduced from the particle output position (i.e., by retrodiction). From Equations (3.6) it follows that

$$\langle AB\rangle = \cos(\theta_B-\theta_A), \quad \langle AB'\rangle = \cos(\theta_{B'}-\theta_A), \qquad (3.7)$$

which are just the negatives of correlations for two particles in a Bell experiment. However, the situation may now be interpreted as measurement *A* correlated with counterfactual *alternatives B* and *B'* computationally predicted for the same kind of random event as Bell envisioned, with both values determined by each (hidden) random variable outcome as shown in Fig. 2.

The crux of the matter is that one can now compute $\langle BB'\rangle$ rather than assume that it has the same form as the correlations in (3.7). The correlation of measurements at alternative settings *B* and *B'* can be computed from Fig. (2). For *A* = 1, the value of

$$\langle BB'|A=1\rangle = (-1)(-1)\sin^2\frac{\theta_B-\theta_A}{2}+(1)(-1)\left(\cos^2\frac{\theta_B-\theta_A}{2}-\cos^2\frac{\theta_{B'}-\theta_A}{2}\right)$$
$$+(1)(1)\cos^2\frac{\theta_{B'}-\theta_A}{2}=1-\cos(\theta_B-\theta_A)+\cos(\theta_{B'}-\theta_A). \qquad (3.8)$$

Since $\langle BB'|A=-1\rangle = \langle BB'|A=1\rangle$, and $P(A=1)=P(A=-1)=1/2$,

$$\langle BB'\rangle = \langle BB'|A=1\rangle P(A=1)+\langle BB'|A=-1\rangle P(A=-1)$$
$$= 1+\cos(\theta_{B'}-\theta_A)-\cos(\theta_B-\theta_A). \qquad (3.9)$$

Bell's inequality is now satisfied by cross-correlations (3.7) and (3.9). No nonlocal information is used by the algorithm to determine *B* and *B'* since only a single particle is considered whereas if the correlations are predicted for measurements on two particles on opposite sides of a Bell apparatus as in [8], the sign of $\langle AB\rangle$ and $\langle AB'\rangle$ are reversed, and the algorithm requires that *B* and *B'* know *A*'s outcome and setting angle, $\theta_A$.

The point of this example is that, in the present case, the same sets of Bell cosine correlations (up to a minus sign that does not effect validity of (2.4)) are associated with the use of local information as are associated with nonlocal information in the case given previously in [8]. However, it is the use of *incorrect assumptions regarding the form of the correlations* for the non-commuting counterfactuals that causes violation of the Bell inequality. Correcting them results in satisfaction of the inequality.

## 4. Discussion and Conclusion

In generating his theorem, Bell misjudged the effect of one of the distinguishing features of quantum formalism; the non-commutation of some observables. The author speculates that this may have resulted from a belief that non-commutation had no place in any theory proposed as an alternative to quantum mechanics. However, as pointed out in Sec. (1a) (and with physical examples given in [1]), non-commutation is frequently encountered in the macroscopic world, although it has not been incorporated into an all-encompassing formalism.

Three key results have been demonstrated in the current paper: (1) The Bell inequality is identically satisfied by the simultaneous cross-correlations for any three data sets. (2) Recent Bell experiments may be adapted to the production of two real measurements and a predicted corresponding counterfactual (in spite of the assumption of non-locality) to produce three data sets. (3) Essentially the same correlations may be predicted for a Bell-measurement pair plus counterfactual, as between two sequential measurements plus a counterfactual on one particle, the latter not involving an implication of non-locality. When the counterfactuals in these examples are computed and correlated using quantum probabilities consistent with non-commutation, both sets of correlations satisfy the Bell inequality. The fact that the resulting correlations are not second-order-stationary, contrary to Bell's assumption, should not be surprising given the qualitative differences between commuting and non-commuting observables. Finally, there is no longer reason to appeal to non-local influences or non-reality to account for a discrepancy between observations and their quantum statistical description. Note, that accounting for the basic facts of a physical situation with the least number of assumptions is a principle goal of science.

It should be stated that while the three correlation Bell inequality has been considered here for simplicity in analyzing Bell's reasoning, the same logical approach may be applied to analysis of the four variable CHSH-Bell inequality [17] since it is identically satisfied under simultaneous cross-correlation of four variables. As above, the non-commutation of variables beyond two while considering two particles implies that the four variables cannot result from a SOS process. However, if the four correlations are computed rather than assumed, they must satisfy the Bell inequality.

Note that Bell inequalities may be applied to all real data as opposed to real and counterfactual data [18]. This requires that the data be properly taken and arranged in sets for SCC after which they satisfy the Bell inequality. The consequences of non-commutation also apply to treatment of the Wigner form of the Bell inequality [19] where the correlations are replaced by probabilities that generate them. The probabilities for the variable pairs involved must be consistent with their commutation and non-commutation as appropriate.

Logical flaws in the Bell theorem lead one to consider whether the entanglement predicted Bell correlation may be derived in other ways. The construction of a local probability model [20] shows that the entanglement derivation of the Bell correlation is not unique. Several other models have been proposed by researchers. A discussion of their relative merits and the possibility of physical models (as opposed to algorithmic probability models) raises questions beyond the scope of this article.

**Aknowledgements**

I would like to thank Armen Gulian and Joe Foreman of the Quantum Studies Group at Chapman University, Burtonsville, MD for useful discussions, and critiques of the material presented here.

**Appendix**

Hidden assumptions in Bell's derivation of the inequality

In Bell's derivation, a set of data pairs originates in Stern-Gerlach spin measurements on entangled particle pairs. (See Figure 1.) (Analogous polarization measurements are carried out on down-converter produced photon pairs [21].) Bell defined hidden-variable based measurement readout functions $A(a = \theta_A, \lambda) = \pm 1$, and $B(b = \theta_B, \lambda) = \pm 1$, to represent measurements carried out on each of the two particles respectively, on opposite sides of a Bell apparatus, using detector angular settings $a$ and $b$. Each new particle pair implies a new value of the random parameter $\lambda$. By assuming $A(a, \lambda) = -B(a, \lambda)$, the entanglement-predicted opposite values of spin are reproduced at equal angular settings, and the universe of possible experimental outcomes is represented by one function: $A(a, \lambda)$.

Bell computed average correlations of the measurement readout functions using a conventional probability density $\rho(\lambda)$ rather than by applying averaging to three finite data sets as in Inequality (2.3), for which the inequality already holds as a result of simultaneous cross-correlation. Thus, using Bell's readout function, $\langle ab \rangle$ of Inequality (2.4) becomes:

$$\langle A(a,\lambda)B(b,\lambda) \rangle = \int A(a,\lambda)B(b,\lambda)\rho(\lambda)d\lambda \ . \tag{A1}$$

To produce an inequality, Bell needed three variables corresponding to three angular settings $a$, $b$, and $c$ of the readout function $A$. Since only one measurement was carried out on each of two particles, a third readout was chosen to be an *alternative* unperformed (counterfactual) measurement. The Bell counterpart to the left side of Inequality (2.4) (before application of the absolute value) is

$$\langle A(a)B(b) \rangle - \langle A(a)B(c) \rangle = \int d\lambda \rho(\lambda)[A(a,\lambda)B(b,\lambda) - A(a,\lambda)B(c,\lambda)] \ . \tag{A2}$$

Using $A(a,\lambda) = -B(a,\lambda)$,

$$-(\langle A(a)A(b) \rangle - \langle A(a)A(c) \rangle) = \int d\lambda \rho(\lambda) A(a,\lambda)A(b,\lambda)[A(b,\lambda))A(c,\lambda) - 1] \tag{A3}$$

Taking absolute values on both sides produces:

$$|\langle A(a)A(b) \rangle - \langle A(a)A(c) \rangle| \le \int d\lambda \rho(\lambda)[1 - A(b,\lambda))A(c,\lambda)] \ ,$$

or

$$|\langle A(a)A(b) \rangle - \langle A(a)A(c) \rangle| \le 1 - \langle A(b)A(c) \rangle \ . \tag{A4}$$

The Bell derivation of Equations (A2-A4) obscures the fact that the inequality results from simultaneous cross-correlation, and further, does not recognize that it holds for finite data sets without ensemble averaging as a consequence of the laws of algebra independently of any statistical properties.

It is important to note [14] that if nonlocal interactions are assumed ab initio between the two sides of a Bell apparatus, $A(a,\lambda)$ must be replaced by $A(a,b,\lambda)$, and a change of $b$ to $b'$ causes the value of $A$ to change. The fundamental algebraic relations of (A2) or (2.1) are then altered, and the inequality no longer holds. This has been a widely used explanation for the violation of the Bell inequality. However, as already indicated above, this situation may be remedied by completing the measurement at $A$ (by lengthening the path on the $B$ side) before the measurement setting at $B$ is decided. $A$ could be measured at an initial setting for $B$ of $b_0$, and this could be randomly changed before the measurement at $B$ or $B'$ occurs. Three data sets would still exist with $B'$ a counterfactual, whether or not there is an influence from $A$ to the $B$-side. The inequality (2.3) would still apply to any resulting three data sets that could in principle be written down, even if the $B'$ data were incorrectly constructed.

The conclusions of the Bell theorem depend on additional *unstated* assumptions applied to Inequality (A4). First, from Bell's statements in [6], he apparently believed that the problem of non-commutation was solved by using a counterfactual value for the third measurement. Second, the *single* process $A(\theta,\lambda)$ that Bell used was assumed to be second order stationary (SOS) in angle [7] in order to yield the correlation $-\cos(\theta_a - \theta_b)$ predicted by quantum mechanics for the *A-B* measurements. The assumption of this functional form for *all* correlations among real and non-commuting counterfactual variables is a pivotal error of the Bell theorem. Stated mathematically, it was assumed that:

$$\langle A(a)A(b) \rangle = -f(\theta_a - \theta_b), \ \langle A(a)A(c) \rangle = -f(\theta_a - \theta_c), \ \langle A(b)A(c) \rangle = -f(\theta_b - \theta_c) \tag{A5.1}$$

with the function $f$ given by

$$f = \cos \ . \tag{A5.2}$$

The correlations now correspond to pairs of measurements on opposite sides of the apparatus, thus reversing the sign of the correlation term on the right-hand side of Inequality (A4).

For each realization of the random process used, one parameter $\lambda$ determines $A(a,\lambda)$ at all $a$. This is consistent with the definition of a SOS process but is inconsistent with a non-commutative process for which each new outcome at a different setting $a$ requires an additional random event, whose probability is conditional on the outcome of the preceding event. *As is shown above, correlations of alternative counterfactuals of such non-commuting operations have a different form from that assumed in Equations (A5.1-A5.2), and the resulting sets of correlations satisfy the Bell inequality.*

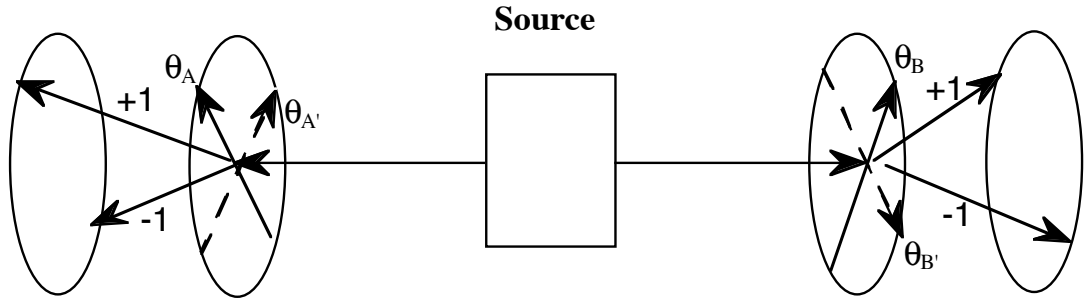

Figure 1. Schematic of Bell experiment in which a source sends two particles to two detectors having angular settings $\theta_A$ and $\theta_B$, and/or counterfactual settings $\theta_{A'}$ and $\theta_{B'}$. While one measurement operation on the A-side, e.g. at setting $\theta_A$, commutes with one on the B-side at $\theta_B$, any additional measurements at either $\theta_{A'}$ or $\theta_{B'}$ are non-commutative with prior measurements at $\theta_A$ and $\theta_B$, respectively.

**Diagram for Hidden Variable Construction**

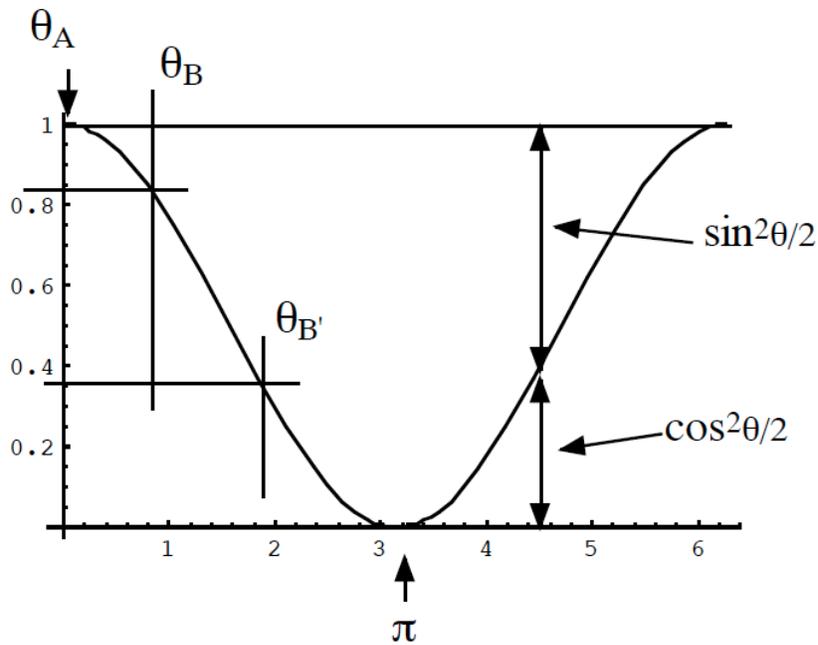

Figure 2. A uniformly distributed random (hidden) variable $\lambda$ may be used to simulate relative probabilities $\cos^2\theta/2$ and $\sin^2\theta/2$ since its values lie above or below the graphed function a corresponding fraction of the time. The angle $\theta$, evaluated at $\theta_B - \theta_A$ and $\theta_{B'} - \theta_A$, specifies points on the curve so that each $\lambda$ determines outcomes at two alternative settings $\theta_B$ and $\theta_{B'}$.